\documentclass[aps,pra,twocolumn,superscriptaddress]{revtex4-1}

\usepackage{graphicx,amsfonts,latexsym,color,dcolumn,bm}
\usepackage{braket}
\usepackage{makeidx}
\usepackage{multirow}
\usepackage[dvipsnames,svgnames,table]{xcolor}
\usepackage{graphicx}
\usepackage{epstopdf}
\usepackage{ulem}
\usepackage{hyperref}
\usepackage{amsmath}
\usepackage{amssymb}

\begin{document}
\newcommand{\be}{\begin{equation}}
\newcommand{\ee}{\end{equation}}
\newcommand{\bea}{\begin{eqnarray}}
\newcommand{\eea}{\end{eqnarray}}
\newcommand{\ad}{a^{\dag}}
\newcommand{\la}{\langle}
\newcommand{\ra}{\rangle}
\newcommand{\om}{\omega}


\title{Induced Coherence, Vacuum Fields, and Complementarity in Biphoton Generation}
\author{A. Heuer}
\affiliation{University of Potsdam, Institute of Physics and Astronomy, Karl-Liebknecht Stra\ss e 24-25,
D-14476 Potsdam}
\author{R. Menzel}
\affiliation{University of Potsdam, Institute of Physics and Astronomy, Karl-Liebknecht Stra\ss e 24-25,
D-14476 Potsdam}
\author{P. W. Milonni}
\affiliation{Theoretical Division, Los Alamos National Laboratory, Los Alamos, New Mexico 87545 USA} 
\affiliation{Department of Physics and Astronomy, University of Rochester, Rochester, NY 14627, USA} 
\date{}

\begin{abstract}
We describe spontaneous parametric down-conversion experiments in which induced coherence across two coupled interferometers results in
high-visibility single-photon interference. Opening additional photon channels allows ``which-path" information and reduces the visibility of the single-photon interference, but results in nearly perfect visibility when photons are counted in coincidence. A simplified theoretical model 
accounts for these complementary observations and attributes them directly to the relations among the vacuum fields at the different crystals.
 
\end{abstract}



\maketitle
Complementarity, often discussed in quantum optics as wave-particle dualism, is a fundamental principle of undiminished interest \cite{aharonov,storey,scully,kocsis}. ``Which-path" information about a single photon is possible only at the cost of decreasing interference fringe visibility $V$. This can be expressed quantitatively by the formula $K^2+V^2\le 1$, where the ``contrast" $K$ is a measure of which-path information \cite{englert}. Recent work \cite{menzel1,menzel2} demonstrated among other things that single-photon interference is determined by the mode function associated with the photon, as quantum field theory predicts \cite{menzel1}. In this paper we describe experiments on spontaneous photon down-conversion (SPDC) demonstrating that which-path information in single-photon interference reduces fringe visibility but results in very high visibility in second-order interference measurements. These experiments may serve to clarify the roles of mode functions and vacuum fields in SPDC, which has for many years been a primary platform for fundamental studies in quantum optics and conceptual foundations of quantum theory.

Although the signal and idler photons in SPDC have no fixed phase relation, the combined entity, or biphoton, carries observable phase information about the pump field \cite{ou,zou}. In particular, by varying the phase delay between the pump fields for two crystals, first-order single-photon interference has been observed \cite{heuer}. In these experiments the idler photon modes i1 and i2 of crystals BBO1 and BBO2 (Figure 1) were aligned and indistinguishable, and interference of the signal photon channels s1 and s2 was observed as a consequence of  induced coherence between BBO1 and BBO2. 

Suppose now that a third pumped crystal (BBO3) is added. If the s1 and s3 modes are aligned and indistinguishable, the possibility of generating a signal photon at BBO1 or at BBO3 might suggest that single-photon interference between the channels s1 and s3 should be observable at detector A behind beam splitter BS1. But we now have ``which-path" information: detection of photons i3 with a detector D behind the beam splitter BS2, for instance, implies that a photon pair (i3,s3) must have been emitted from crystal BBO3. In spite of this
which-path information, however, we observe high-visibility signal-photon interference fringes at A when these signal photons are counted in coincidence with idler photons at D.

The experimental setup employed three BBO crystals for biphoton generation (Figure 1). All three crystals had a length of 4 mm and were cut for type I phase matching. The crystals were arranged sequentially for induced coherence in 
pairs \cite{zou,wang,heuer}. The idler channel i1 of crystal BBO1 was matched to the idler channel i2 of crystal BBO2 and the signal channel s1 of BBO1 was matched to the signal channel s3 of BBO3. Induced coherence and phase memory therefore occur between crystals BBO1 and BBO2 and between crystals BBO1 and BBO3, as described previously \cite{heuer}. The three pump beams were obtained from a laser (Genesis, Coherent) which emitted an almost diffraction-limited cw field at 355 nm with a bandwidth of about 45 GHz and a coherence length of about 1.4 mm, and were synchronized with delay lines in front of the three crystals. Spectral filters were used such that signal and idler fields had wavelengths of 808 nm and 632 nm, respectively. For varying the length of the signal and the idler single-photon interferometers the 100\% reflecting mirrors could be moved, resulting in a phase shift for the signal photons (upper mirror in Figure 1) and for the idler photons (lower mirror in Figure 1). The photons were detected with photodiodes (SPCM-AQRH-13, Perkin Elmer) at the positions A and D. The bandwidth was about 1 nm (FWHM) for the idler photons and 2 nm for the signal photons. The pump powers at all three crystal were 30 mW. With high probability just one photon pair was in the apparatus during the measurement interval 2 ns. 

\begin{figure}[h!]
\includegraphics[width=8 cm]{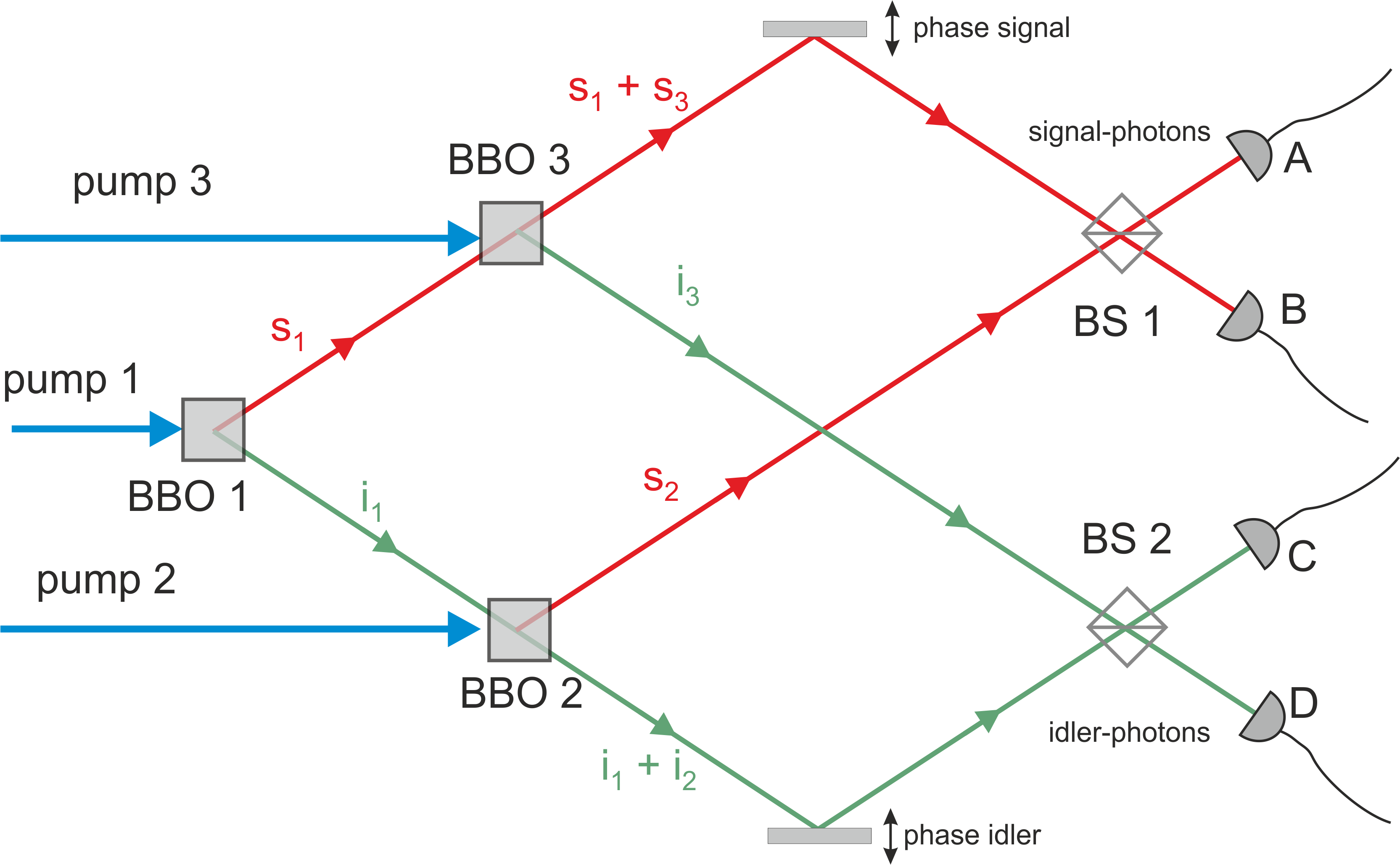}
\caption{Experimental setup employing three synchronously pumped SPDC crystals (BBO1, BBO2, BBO3) and induced coherence in the measurement of single-photon and biphoton interference.}
\label{figure1}
\end{figure}

In the first set of experiments only crystals BBO1 and BBO2 were pumped and the idler modes i1 and i2 were aligned to be indistinguishable, and consequently the signal photons from these crystals showed first-order interference at the detector A (Figure 2). The interference fringes clearly show a period of 808 nm, the signal wavelength. The measured fringe visibility $V=80\%$, even without any correction for  background photons. The same fringe pattern, with a visibility $V=96\%$, is obtained when the signal photons are measured in coincidence with the photons in the idler channel (Figure 3), indicating almost perfect induced coherence between BBO1 and BBO2. Similar results are obtained when measurements are made while pumping only BBO1 and BBO3. To check for coherence between BBO2 and BBO3, only these two crystals were pumped and signal and idler photons were detected in coincidence between the detectors A and D. The interference fringe visibility is again high, $V=88\%$.

\begin{figure}[h!]
\includegraphics[width=7 cm]{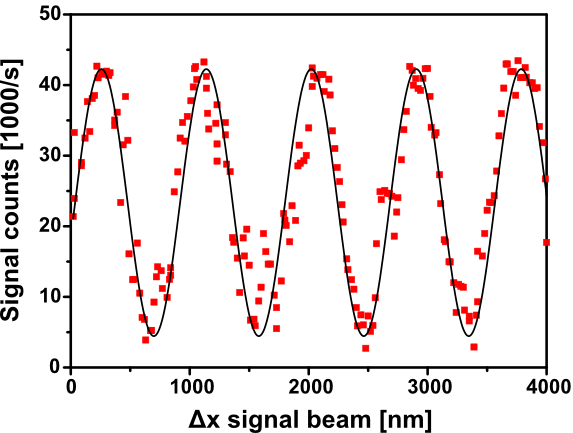}
\caption{First-order interference at detector A of signal photons emitted from crystals BBO1 and BBO2.}
\label{figure2}
\end{figure}

\begin{figure}[h!]
\includegraphics[width=7 cm]{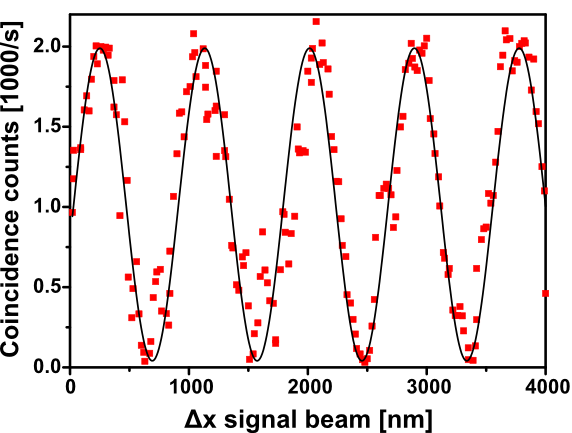}
\caption{Second-order interference at detector A of signal photons emitted from BBO1 and BBO2 in coincidence with the correlated idler photons at detector D.}
\label{figure3}
\end{figure}

In the next set of experiments all three pump channels were opened. The measurement of signal photons with detector A shows the same fringe structure as in Figure 1, but now an incoherent background appears (Figure 4). Thus, although we know from the measurements above and the results of Reference \cite{heuer} that the signal mode in the channel s3 of crystal BBO3 matches almost perfectly the mode of channel s1, BBO3 adds an incoherent background; this background background is found to have exactly the same strength as the count rate of signal photons from crystal BBO3, and reduces the fringe visibility to $V=53\%$. 

Interference was then observed in coincidence at detectors A and D when all three crystals were pumped, but in this measurement the delay lines for the signal and idler photons were continuously moved with constant velocities of 20 nm/s and 10 nm/s, respectively. The result of this measurement as a function of time is shown in Figure 5, and it is seen that the coincidence rate shows almost zero values at some times, i.e., there are certain combinations of signal and idler delays  that result in almost perfect visibility in the coincidence measurement, consistent with the assumption that all photon channels are constituted from coherent photon modes, allowing strong interference.

\begin{figure}[t]
\includegraphics[width=7 cm]{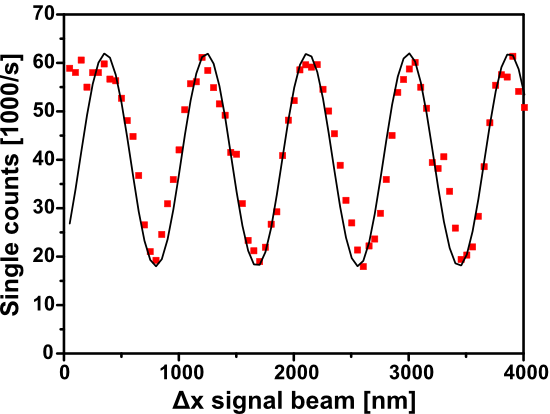}
\caption{First-order interference at detector A of the signal photons from crystals BBO1, BBO2 and BBO3. Besides the fringes there is an incoherent background. The count rate of this background is equal to the signal photon emission rate of crystal BBO3.}
\label{figure4}
\end{figure}

\begin{figure}[t]
\includegraphics[width=7 cm]{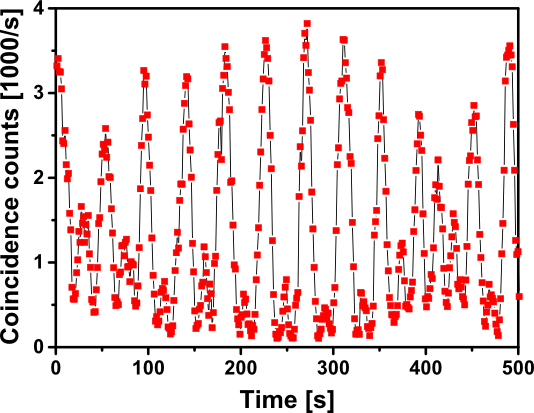}
\caption{Second-order interference of the signal photons at detector A in coincidence with the correlated idler photons at detector D when all three crystals are pumped and the signal and idler phase delays are varied with different speeds. The very high contrast for certain phase relations indicates a very high degree of induced coherence and phase memory.}
\label{figure5}
\end{figure}

All these results can be explained with a simplified model for SPDC based on an effective Hamiltonian with couplings $a_P\ad_S\ad_I$ and 
$\ad_Pa_Sa_I$, where $a_P,a_S$, and $a_I$ ($\ad_P,\ad_S$, and $\ad_I$) are photon annihilation (creation) operators for the pump, signal, and idler fields, respectively. This describes the annihilation (creation) of a pump photon and the simultaneous creation (annihilation) of signal and idler photons. We assume perfect phase matching and that the signal and idler photons generated at each crystal are excitations of single field modes of frequency $\om_S$ and $\om_I$, respectively. The pump is treated as an undepleted, classical field. From the Heisenberg equations of motion for the field operators we write the photon annihilation part of the total electric field operator for the signal field at detector A as
\bea
E_A^{(+)}&=&\big[a_{s20}+ia_{s10}e^{i\phi_S}+iC_1e^{i\phi_S}\ad_{i10}+C_2\ad_{i10}\nonumber\\
&&\mbox{}+iC_3e^{i\phi_S}\ad_{i30}\big]e^{-i\om_St},
\label{th1}
\eea
ignoring an irrelevant multiplicative constant. The first term on the right is the vacuum contribution to the field
mode s2 at detector A, and the second term is the vacuum contribution to the mode s1, including explicitly the phase shift $\phi_S$ and a factor $i$ associated with reflection at the beam splitter BS1 in Figure 1; these terms contribute independently of any pump fields. The third term is the contribution to the signal field at detector A that results from the mixing of the pump field at BBO1 with the vacuum idler field in mode i1. The constant $C_1$ is proportional to the electric field strength of the pump at BBO1, and again we include the phase factor due to the upper mirror. The fourth term is the contribution resulting from the mixing in BBO2 of the vacuum idler field in mode i1 with the pump field incident on BBO2, while the last term comes from the mixing of the pump at BBO3 with the vacuum idler field in mode s3, including the effect of the upper mirror. $C_2$ and $C_3$ are proportional to the electric field strengths of the pump fields at crystals BBO2 and BBO3, respectively. The photon creation part of the total signal field at A is $E_A^{(-)}=[E_A^{(+)}]^{\dag}$.

In writing $E_A^{(+)}$ we have assumed that the idler mode that mixes with the pump field at BBO1 to generate a signal photon in mode s1 is the same idler mode that mixes with the pump at BBO2 to generate a signal photon in mode s2. Therefore $\ad_{i10}$ appears in both the third and fourth terms of
Eq. (\ref{th1}). This is consistent with the assumption that stimulated down-conversion is
negligible, and will be seen in our model to be the physical origin of the phase coherence of the signal fields s1 and s2. 

In similar fashion and notation we express the photon annihilation part of the total electric field operator for the idler field at detector D in Figure 1 as
\bea
E_D^{(+)}&=&\big[a_{i30}+ia_{i10}e^{i\phi_I}+iC_1e^{i\phi_I}\ad_{s10}+iC_2e^{i\phi_I}\ad_{s20}\nonumber\\
&&\mbox{}+C_3\ad_{s10}\big]e^{-i\om_It}.
\label{th2}
\eea
As in our assumption in expression (\ref{th1}) that the modes i1 and i2 are identical, we have assumed the identity of the modes s1 and s3. This is the physical origin in our model of the phase coherence of the idler fields i1 and i3. 

From $E_A^{(+)}$ and $E_D^{(+)}$ we obtain the normally ordered correlation functions determining the counting rates measured by ideal photon detectors A and D. These rates are calculated by taking vacuum expectation values, since we are dealing with spontaneous down-conversion in which all signal and idler modes are initially unoccupied. Consistent with the assumption of low conversion efficiency, we retain only the lowest-order terms in the $C$'s.  The signal photon counting rate at A, for example, is proportional to 
\bea
R_{SA}&=&\la E_A^{(-)}E_A^{(+)}\ra\nonumber\\
&=&|C_2+iC_1e^{i\phi_S}|^2\la a_{i10}\ad_{i10}\ra+|C_3|^2\la a_{i30}\ad_{i30}\ra\nonumber\\
&=&|C_2+iC_1e^{i\phi_S}|^2+|C_3|^2.
\label{th3}
\eea
The coincidence counting rate $R_{SA,ID}$ for signal photons at A and idler photons at D, similarly, is proportional to 
\bea
R_{SA,ID}&=&\la E_A^{(-)}E_D^{(-)}E_D^{(+)}E_A^{(+)}\ra\nonumber\\
&=&|iC_1e^{i\phi_S}+C_2|^2\la a_{i10}\ad_{i10}a_{i10}\ad_{i10}\ra\nonumber\\
&&\mbox{}+|C_3|^2\la a_{i30}\ad_{i30}a_{i30}\ad_{i30}\ra\nonumber\\
&&\mbox{}+2{\rm Re}\Big\{C_3^*[iC_1e^{i\phi_S}+C_2]e^{-i(\phi_S-\phi_I)}\Big\}\nonumber\\
&&\mbox{} \ \ \ \ \ \times\la a_{i30}\ad_{i30}a_{i10}\ad_{i10}\ra\nonumber\\
&=&\big|iC_1e^{i(\phi_S+\phi_I)}+C_2e^{i\phi_I}+C_3e^{i\phi_S}\big|^2.
\label{th4}
\eea
If only BBO1 and BBO2 are pumped, for instance, $C_3=0$ and $R_{SA}$ exhibits high-visibility interference fringes with periodicity equal to the signal field wavelength (Figure 2). In this case there are two indistinguishable paths to a signal photon count at A: s1 $\rightarrow$ upper mirror $\rightarrow$ A and s2 $\rightarrow$ A. These two paths have probability amplitudes $iC_1e^{i\phi_S}$ and $C_2$, respectively, and combine to give the probability $|C_2+iC_1e^{i\phi_S}|^2$. The fringe visibility in this case is $V=2|C_1C_2|/(|C_1|^2+|C_2|^2)$, while the contrast $K\equiv||C_1|^2-|C_2|^2|/(|C_1|^2+|C_2|^2)$, implying $V^2+K^2=1$. Allowing for some degree of pump incoherence, we obtain more generally the relation $V^2+K^2\le 1$ noted earlier \cite{englert}.
 
If all three crystals are pumped, an additional channel s3 $\rightarrow$ upper mirror $\rightarrow$ A is opened which is distinguishable from s1 $\rightarrow$ upper mirror $\rightarrow$ A and s2 $\rightarrow$ A. In this case it follows 
from (\ref{th3}) that there is superposed on the fringes of Figure 2 an incoherent background count rate equal to the rate of signal photon generation ($|C_3|^2)$ when only BBO3 is pumped (Figure 4). Similar calculations for the idler photon counting rate $R_{ID}$ account for the experimental results for the idler photon counting rate at detector D.

$R_{SA,ID}$, similarly, accounts for the experimental results shown in Figures 3 and 5. If only BBO1 and BBO2 are pumped, the second-order interference of signal photons at A in coincidence with idler photons at D shows the same variation with $\phi_S$ as the first-order interference of signal photons from BBO1 and BBO2 at A, and the fringe visibility is high (Figure 3). When all three crystals are pumped, there are three indistinguishable paths by which a signal photon can be counted at A and an idler photon counted at D: (i) s1 $\rightarrow$ upper mirror $\rightarrow$ A, i1 $\rightarrow$ lower mirror $\rightarrow$ D; (ii) s2 $\rightarrow$ A, i2 $\rightarrow$ lower mirror $\rightarrow$ D; and (iii) s3 $\rightarrow$ upper mirror $\rightarrow$ A, i3 $\rightarrow$ D. The probability amplitudes for these processes are (i) $i^2C_1e^{i\phi_S}e^{i\phi_S}$; (ii) $iC_2e^{i\phi_I}$; and (iii) $iC_3e^{i\phi_S}$, resulting in a probability proportional to $|i^2C_1e^{i\phi_S}e^{i\phi_S}+iC_2e^{i\phi_I}+iC_3e^{i\phi_S}|^2$, in agreement with equation (\ref{th4}). $R_{SA,ID}$ in this case exhibits very high fringe visibility for certain relations between the phases $\phi_S$ and $\phi_I$, as found in the experiments (Figure 5). 

In summary, induced coherence between two crystals BBO1 and BBO2, and between BBO1 and BBO3, results in high-visibility interference in photon counting of signal (or idler) photons (Figure 2) as well as in coincidence counting of signal and idler photons (Figure 3). This is consistent with the matching of the idler modes i1 and i2 and the matching of the signal modes s1 and s3. The addition of a third pumped crystal BBO3, however, introduces ``which-path" information and a substantial decrease in the visibility of {\sl single-photon} interference fringes (Figure 4), in spite of the indistinguishability of the modes s1 and s3 for the signal photons and of the modes i1 and i2 for the idler photons. At the same time the
third pumped crystal results in nearly perfect visibility when photons are counted in coincidence (Figure 5).

These results also demonstrate the physical significance of the vacuum fields taking part in the biphoton generation at the different crystals. In this {\sl spontaneous} biphoton generation the vacuum idler fields incident on BBO1 and BBO2 are identical, resulting in coherence (interference) between the s1 and s2 modes. But different (and uncorrelated) vacuum idler fields are incident on BBO1 and BBO3, and therefore, in contrast to stimulated coherence measurements, there is no interference between the s1 and s3 modes, in spite of their being spatially indistinguishable. The fact that the vacuum idler fields incident on BBO1 and BBO3 are uncorrelated implies in our model that $\la a_{i10}\ad_{i30}\ra=0$, and that the contributions $\la a_{i10}\ad_{i10}\ra$ and $\la a_{i30}\ad_{i30}\ra$ of these fields to the signal single-photon interference appear additively in Eq. (3); this addition is responsible for the incoherent background of Figure 4. But pumping the third crystal (BBO3), while introducing a ``which-way" channel in single-photon counting experiments,  results in very high fringe visibility in the complementary experiment in which signal and idler photons are counted in coincidence (Figure 5). We can interpret this, as above, to the opening by the third crystal of additional, indistinguishable ways (paths (i), (ii), and (iii) above) in which signal and and idler photons can be counted {\sl in coincidence}. In terms of the vacuum fields, we can also attribute this to the fact that the vacuum idler contributions in the three-crystal coincidence measurements are of the form 
$\la a_{i30}\ad_{i30}a_{i10}\ad_{i10}\ra$, and are nonzero regardless of the fact that the i1 and i3 vacuum fields are uncorrelated.

\end{document}